# High Thermal Conductivity in Wafer-Scale Cubic Silicon Carbide Crystals

Zhe Cheng[1,*], Jianbo Liang[2,*], Keisuke Kawamura[3], Hidetoshi Asamura[4], Hiroki Uratani[3], Samuel Graham[5], Yutaka Ohno[6], Yasuyoshi Nagai[6], Naoteru Shigekawa[2], David G. Cahill[1,*]

[1] Department of Materials Science and Engineering and Materials Research Laboratory, University of Illinois at Urbana-Champaign, Urbana, IL 61801, United States.
[2] Department of Electronic Information Systems, Osaka Metropolitan University, Sugimoto 3-3-138, Sumiyoshi, Osaka 558-8585, Japan.
[3] The SIC Division, Air Water Inc. 2290-1 Takibe, Toyoshina Azumino, Nagano 399-8204, Japan.
4 Specialty Materials Department, Electronics Unit, Air Water Inc. 4007-3 Yamato, Azusagawa, Nagano 390-1701, Japan.
[5] George W. Woodruff School of Mechanical Engineering, Georgia Institute of Technology, Atlanta, GA 30332, USA.
[6] Institute for Materials Research, Tohoku University, 2145-2 Narita, Oarai, Ibaraki 311-1313, Japan.

[*]Corresponding authors: zcheng18@illinois.edu; liang@omu.ac.jp; d-cahill@illinois.edu.

**ABSTRACT**

High thermal conductivity electronic materials are critical components for high-performance electronic and photonic devices as either active functional materials or thermal management materials. We report an isotropic high thermal conductivity over 500 W m$^{-1}$K$^{-1}$ at room temperature in high-quality wafer-scale cubic silicon carbide (3C-SiC) crystals, which is the second highest among large crystals (only surpassed by diamond). Furthermore, the corresponding 3C-SiC thin films are found to have record-high in-plane and cross-plane thermal conductivity, even higher than diamond thin films with equivalent thicknesses. Our results resolve a long-lasting puzzle that the literature values of thermal conductivity for 3C-SiC are perplexingly lower than the structurally more complex 6H-SiC. Further analysis reveals that the observed high thermal conductivity in this work arises from the high purity and high crystal quality of 3C-SiC crystals which excludes the exceptionally strong defect-phonon scatterings in 3C-SiC. Moreover, by integrating 3C-SiC with other semiconductors by epitaxial growth, we show that the measured 3C-SiC-Si TBC is among the highest for semiconductor interfaces. These findings not only provide insights for fundamental phonon transport mechanisms, also suggest that 3C-SiC may constitute an excellent wide-bandgap semiconductor for applications of power electronics as either active components or substrates.

# INTRODUCTION

Silicon carbide (SiC) plays a fundamental role in many emerging technologies such as power electronics, optoelectronics, and quantum computing.[1-4] SiC based power devices can lead a revolution in power electronics to replace Si-based technology due to its fast switching speeds, low losses, and high blocking voltage.[5] In power electronics and optoelectronics, the high localized heat flux leads to overheating of devices.[6,7] The increased device temperature degrades their performance and reliability, making thermal management a grand challenge.[6,8] High thermal conductivity ($\kappa$) is critical in thermal management design of these electronics and optoelectronics, especially for high-power devices.[9,10]

Current high $\kappa$ electronic materials such as hexagonal SiC and AlN have room-temperature c-axis $\kappa$ of 324 W m$^{-1}$K$^{-1}$ for 6H-SiC, 348 W m$^{-1}$K$^{-1}$ for 4H-SiC, and 321 W m$^{-1}$K$^{-1}$ for AlN, which are lower than metals such as silver and copper (~400 W m$^{-1}$K$^{-1}$).[11,12] It is notable that the widely-used high $\kappa$ value (490 W m$^{-1}$K$^{-1}$) for 6H-SiC is from Slack's measurements back to 1964 with a steady-state technique.[13,14] Recent more advanced measurements based on time-domain thermoreflectance (TDTR) narrowed down the errors and corrected this value to ~320 W m$^{-1}$K$^{-1}$ for 6H-SiC,[11,15,16] which is consistent with first-principles calculations of perfect single crystal 6H-SiC based on density functional theory (DFT).[17]

Compared with the extensively studied and widely used hexagonal phase SiC polytypes (6H and 4H), the cubic phase SiC (3C) is much less well understood even though it potentially has the best electronic properties and much higher $\kappa$.[1,5] The metal oxide semiconductor field effect transistor (MOSFET) based on 3C-SiC has the highest channel mobility ever presented on any SiC polytype, which produces a large reduction in the power consumption of power switching devices.[5] A long-lasting puzzle about the κ of 3C-SiC is that the measured value in the literature is perplexingly lower than that of the structurally more complex 6H phase.[1] This contradicts theoretical calculations that structural complexity and κ are inversely correlated.[17] To explain the abnormally low κ of 3C-SiC in the literature, A. Katre, *et al.* attributed the low κ to exceptionally strong boron defect-phonon scatterings in 3C-SiC, which are even stronger than vacancies.[1] 0.1% boron creates a factor of 2 decrease in κ while the same reduction is created by 2% substitutional nitrogen.[1] However, experimental validation is still lacking partly due to the challenges in growing high-quality crystals.[5,18]

The potentially high κ of 3C-SiC not only facilitates applications which use 3C-SiC as active electronic materials, but also enables 3C-SiC to be a thermal management material which cools devices made of other semiconductors. In terms of thermal management materials, diamond has the highest isotropic κ among all bulk materials but is limited by its high cost, small wafer size, and difficulty in heterogeneous integration with other semiconductors with high thermal boundary conductance (TBC).[10,19,20] Graphite has extremely strong intrinsic anisotropy in κ due to weak cross-plane van der Waals bonding.[21] For other carbon-based nanomaterials such as graphene and carbon nanotubes, their κ decrease significantly when assembling together or with other materials.[7] Recently, great progress has been achieved in the discovery of isotropic high κ in high-purity boron-based crystals, such as cubic BAs,[22-24] natural and isotope-enriched cubic BN,[25] natural and isotope-enriched cubic BP,[25-27] but all the crystal sizes are millimeter-scale or smaller. The technical difficulties in growth of high-purity large crystals prevent these high κ thermal management materials from scalable applications. Further heterogeneous integration of these high κ thermal management materials with other semiconductors with high TBC is also challenging.[28,29]

Here, we report an isotropic high κ over 500 W m$^{-1}$K$^{-1}$ at room temperature in a high-purity wafer-scale free-standing 3C-SiC bulk crystal grown by low-temperature chemical vapor deposition (LT-CVD), which agrees well with the first-principle predicted intrinsic κ of perfect single-crystal 3C-SiC. Moreover, 3C-SiC can be heterogeneously integrated with Si and AlN by epitaxial growth. The in-plane and cross-plane κ of corresponding 3C-SiC thin films are measured by beam-offset time-domain thermoreflectance (BO-TDTR). Further structural analysis such as Raman spectroscopy, X-ray diffraction (XRD), high-resolution transmission electron microscopy (HR-TEM), electron backscatter diffraction (EBSD), and second ion mass spectroscopy (SIMS) are performed to understand the structure-κ relation. Additionally, the TBC of 3C-SiC epitaxial interfaces with Si and AlN are studied by TDTR.

## RESULTS

3C-SiC has a less complex crystal structure than 6H-SiC (Fig. 1a). Higher κ than 6H phase is predicted for 3C-SiC single crystal.[1] We synthesize a free-standing 3C-SiC wafer (Fig. 1b) by growing 3C-SiC on a silicon substrate at 1000 °C in a customized reactor and then etching away the Si substrate. More details about growth techniques can be found in Methods section and Supplementary Information (SI). Peaks (795 cm$^{-1}$ for TO and 969 cm$^{-1}$ for LO) in Raman spectrum measured on the 3C-SiC crystal (Fig. 1c) agree well with the Raman peaks of 3C-SiC in the literature (796 cm$^{-1}$ for TO and 970 cm$^{-1}$ for LO).[30] Fig. 1d shows rocking curve of the X-ray diffraction of the 3C-SiC crystal. The full width at half magnitude (FWHM) of the (111) peak is 158 arcsec, showing the high crystal quality of the 3C-SiC crystal. To probe the crystal structure of the 3C-SiC, we obtained an annular dark field scanning transmission electron microscopy (STEM) image (Fig. 1e) with atomically resolved lattices. The Fast Fourier transform of the STEM image is shown in the inset of Fig. 1e. Fig. 1f shows the selected area electron diffraction (SAED) pattern in a TEM, further confirming the SiC crystal is the cubic phase. More details about Raman measurements, STEM, and SAED can be found in Methods section and SI. The density of stacking faults of the growth surface is low (about 1000 cm$^{-2}$). We performed electron backscatter diffraction (EBSD) measurements on the freestanding bulk 3C-SiC to figure out the crystal orientation. The EBSD data of both the face close to Si substrate and the growth face shows single orientation (111) over the entire scanned area (2.4 mm × 0.8 mm). More details can be found in the SI. To figure out the impurity concentrations in 3C-SiC, secondary ion mass spectrometry (SIMS) was used to measure the concentrations of boron, nitrogen, and oxygen. The concentrations measured from the growth face are 6.6×10$^{17}$ atoms cm$^{-3}$ for oxygen and 5.8×10$^{15}$ atoms cm$^{-3}$ for nitrogen. The concentrations measured from the carbon face (adjacent to the Si substrate before etching away Si) are 2.3×10$^{18}$ atoms cm$^{-3}$ for oxygen and 1.4×10$^{16}$ atoms cm$^{-3}$ for nitrogen. The concentrations of boron impurity are too low to be detectable for SIMS measurements from both faces. The measured low concentrations of impurities further confirm the high quality of the 3C-SiC crystals in this work and high κ is expected.[1]

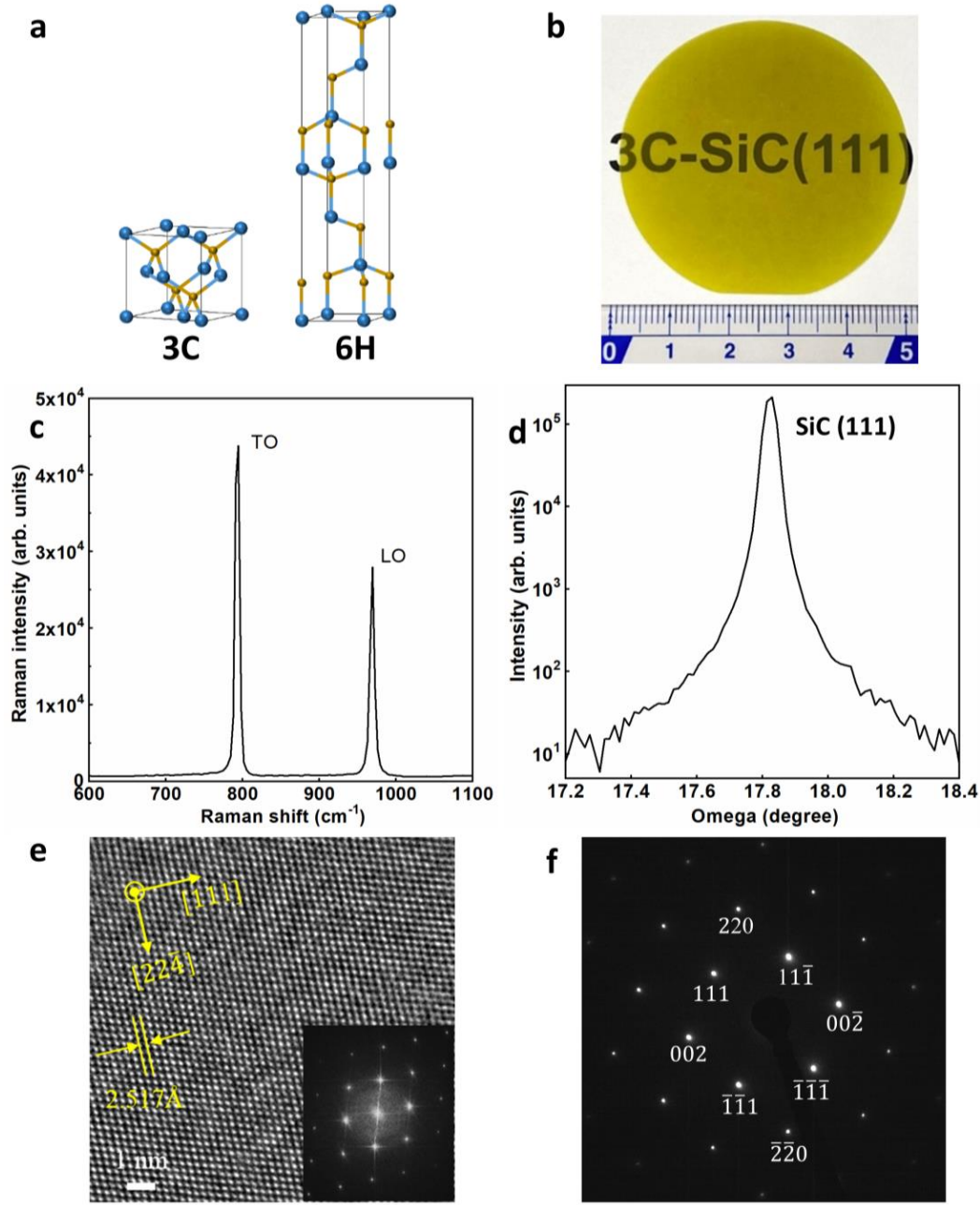

**Fig. 1. Wafer-scale free-standing 3C-SiC bulk crystals. a** Atomic structure of 3C-SiC and 6H-SiC. **b** Picture of a 3C-SiC 2-inch wafer. The unit of the rule is cm. **c** Raman spectrum of 3C-SiC crystal. **d** X-ray diffraction (XRD) of 3C-SiC. **e** High-resolution STEM image of 3C-SiC taken along the [$\bar{1}$10] zone axis. The inset: Fast Fourier transform of the STEM image. **f** Selected area electron diffraction pattern of 3C-SiC taken in the [$\bar{1}$10] zone axis.

We performed TDTR measurements on the free-standing 3C-SiC bulk crystal to obtain its thermal conductivity. Figure 2a shows an example of the TDTR ratio data (circles) and model fitting (solid line) for the bulk 3C-SiC sample with 5X objective and 9.3 MHz modulation frequency. The dash lines are model curves using κ 10% larger or 10% smaller than the best-fit κ to illustrate the measurement sensitivity. More details about the TDTR measurements can be found in the Methods section and SI. To evaluate the effect of ballistic thermal transport on TDTR measurements of high κ samples, we did multiple TDTR measurements with different spot sizes (10.7 μm for 5X objective, 5.5 μm for 10X objective, and 2.7μm for 20X objective) and different modulation frequencies (1.9-9.3 MHz). We observed weak dependence of measured κ on the used modulation frequency (Fig. 2b) while strong reduction in the measured κ for 20X compared to 5X and 10X (Fig. 2b). This reduction is due to the ballistic thermal transport in the sample and the mismatch in the distributions of phonons that carry heat across the metal transducer-sample interface and in the sample.[31] We used 9.3 MHz and 5X objective for the rest measurements on the κ of bulk 3C-SiC (Fig. 2c and Fig. 3).

The measured κ of 3C-SiC at room temperature is compared with other high κ crystals as a function of wafer size (Fig. 2c).[11,12,15,16,20,22,25,26] The recently reported boron-based crystals have high κ but the achievable crystal sizes are millimeter-scale or smaller due to the limits of growth techniques or thermodynamics. The small crystal sizes prevent them from scalable thermal management applications for electronics and optoelectronics. Single crystal diamond has a larger wafer size, up to 1 inch, but the wide-range adoptions are limited by the high cost and difficulty in heterogeneous integration with other semiconductors.[19,20,29] Because of the large lattice mismatch with other semiconductors such as Si and GaN, heterogeneous epitaxial growth of single crystal diamond is challenging.[29] Current chemical vapor deposited (CVD) polycrystalline diamond results in significantly reduced and anisotropic κ.[32,33]

It is notable that the 3C-SiC wafer reported in this work can reach up to 6-inch in size with an isotropic high κ of over 500 W m$^{-1}$K$^{-1}$. The measured κ of 3C-SiC is higher than all metals and the second highest among all large crystals (only surpassed by single crystal diamond). The κ of 3C-SiC at room temperature is ~ 50% higher than the c-axis κ of 6H-SiC and AlN, ~ 40% higher than the c-axis κ of 4H-SiC.

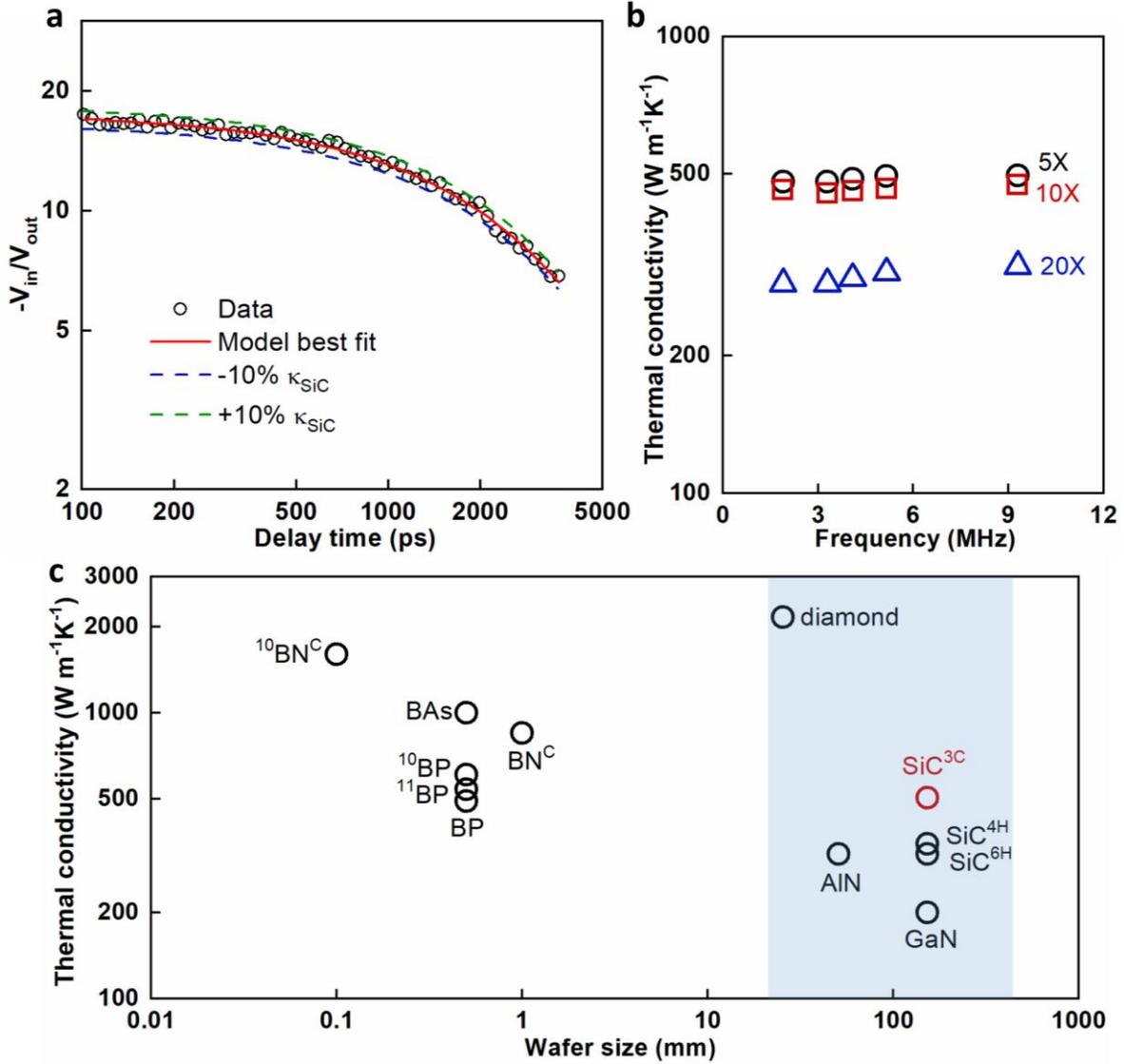

**Fig. 2. High κ of 3C-SiC bulk crystals near room temperature. a** TDTR ratio data (circles) and model fitting (solid line) for 3C-SiC sample. The dash lines are model curves using κ 10% larger or 10% smaller than the best-fit κ to illustrate the measurement sensitivity. **b** Dependence of modulation frequency and laser spot size on the measured κ of 3C-SiC near room temperature. **c** The measured κ of 3C-SiC at room temperature is compared with other high κ crystals as a function of wafer size.[11,12,15,16,20,22,25,26]

We further measured the κ of bulk 3C-SiC crystal at high temperatures. The measured temperature dependent κ of bulk 3C-SiC is compared with previously measured κ values in the literature, κ values of perfect single crystal predicted by DFT, and that of other high κ crystals (See Fig. 3a and 3b). The measured κ agrees excellently with DFT-calculated κ of perfect single crystal 3C-SiC at all measured temperatures. The measured κ in this work is more than 50% higher than the literature values of 3C-SiC at room temperature, and surpasses that of the structurally more complex 6H-SiC. These results are consistent with the theoretical calculations that structural complexity and κ are inversely related.[17] The measured high κ resolves a long-lasting puzzle about the abnormally low κ values in the literature which was attributed to the extrinsic defect-phonon scatterings in 3C-SiC.[1] Boron defects in 3C-SiC cause exceptionally strong phonon scatterings which results from the resonant phonon scattering around the boron impurity.[1] The measured boron impurity concentration is negligible in our 3C-SiC crystals according to the SIMS measurements. The rocking curve of XRD measurements shows a full-width at half magnitude (FWHM) of 158 arcsec. The high-purity and good crystal quality of our 3C-SiC crystals results in the observed high κ, which validates the theory proposed in the literature.[1]

We also compare the measured temperature dependent κ of bulk 3C-SiC crystals with that of AlN, 6H-SiC, and GaN. We include both the in-plane κ and cross-plane κ of 6H-SiC since the κ of 6H-SiC is anisotropic. The DFT-calculated κ values of perfect single crystals agree well with the measured κ values and both are proportional to the inverse of temperature due to the dominant phonon-phonon scatterings in these crystals at high temperatures. The measured κ values of 3C-SiC are 2.5 times as high as that of GaN, making 3C-SiC a potential candidate as substrates of GaN-based power electronics. The high κ of 3C-SiC will motivate the study of power electronics which use 3C-SiC as active device material as a more advanced addition to currently wide-adopted 4H-SiC and 6H-SiC.

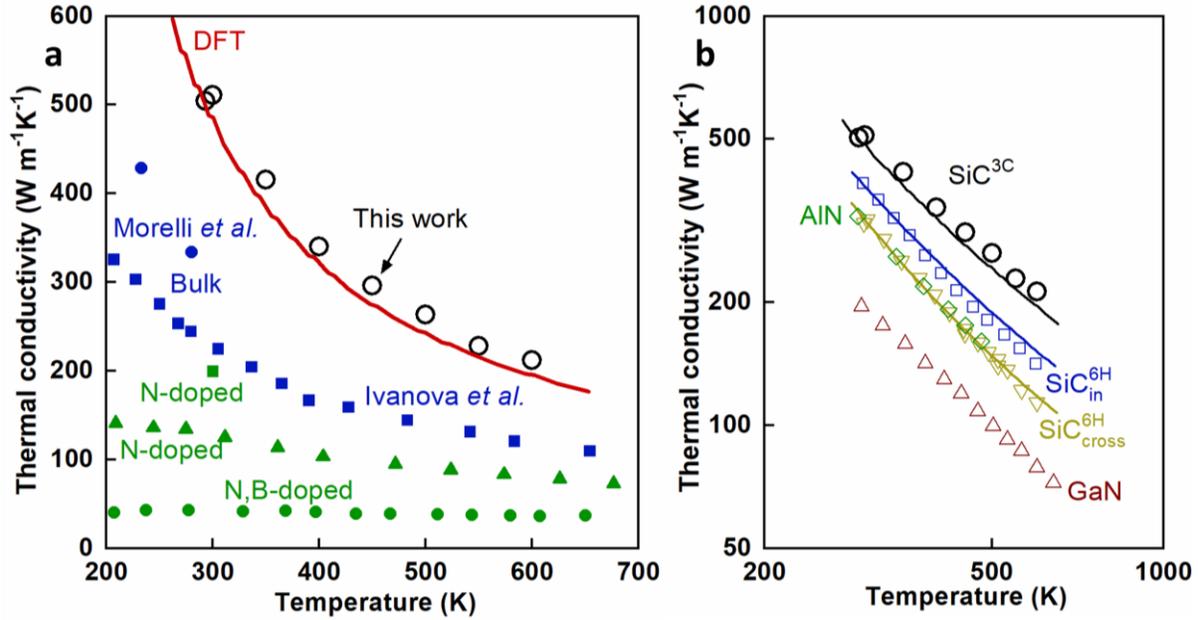

**Fig. 3. Temperature dependent κ of bulk 3C-SiC crystals. a** Comparison of the measured κ in this work with previous measured κ in the literature.[1,34,35] The κ value (red line) predicted by density functional theory (DFT) is also included.[1,36] **b** Comparison of temperature dependent κ of 3C-SiC with c-axis κ of bulk 6H-SiC, AlN, and GaN.[1,11,12] The symbols are experimentally measured values while the lines are DFT-calculated values of perfect single crystals.[1,11] We include both the cross-plane κ and the in-plane κ of 6H-SiC since its κ is anisotropic.

We performed beam-offset time-domain thermoreflectance (BO-TDTR) on 3C-SiC thin films grown on Si substrates to obtain the in-plane κ of 3C-SiC films.[37,38] During BO-TDTR measurements, the pump beam is offset relative to the probe beam, as shown in Fig. 4a. An example of the out-of-phase TDTR signal on a 2.52-µm-thick SiC film on Si sample is shown as a function of the beam offset distance. The full width at half magnitude (FWHM) is a measure of the heat spreading capability which is used to fit for the in-plane κ of the 3C-SiC thin film. More details about the BO-TDTR can be found in the Methods section and SI. The measured in-plane κ values of 3C-SiC thin films at room temperature are compared with that of other close-to-isotropic thin films such AlN, diamond, and GaN (see Fig. 4b). Please note that strongly anisotropic materials graphite and h-BN have high in-plane κ values but we do not include them here. The in-plane κ of 3C-SiC thin films show record-high values, even higher than that of diamond thin films with equivalent thickness. We attribute this high in-plane κ values to the high-quality of the 3C-SiC thin films. This high in-plane κ values of 3C-SiC thin films facilitate heat spreading of localized Joule-heating in power electronics.

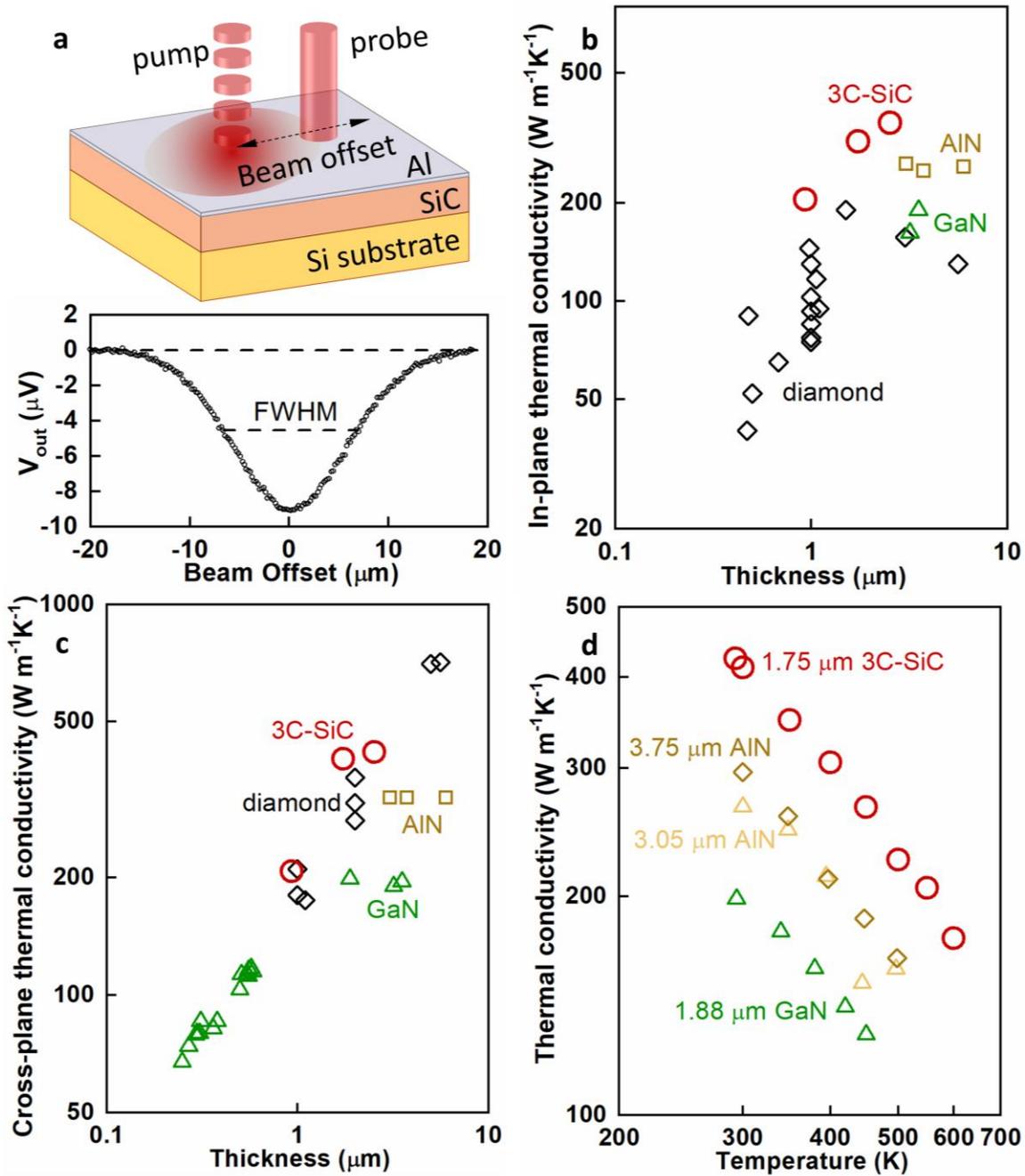

**Fig. 4. High in-plane and cross-plane κ of 3C-SiC thin films. a** Beam-offset TDTR technique for in-plane κ measurements. The out-of-phase TDTR signal on the 2.52 μm SiC on Si sample is shown as a function of beam offset distance. **b** In-plane κ of 3C-SiC thin films. The in-plane κ of other close-to-isotropic high κ thin films are also included for comparison.[32,39-45] **c** Cross-plane κ of 3C-SiC thin films. The cross-plane κ of other high κ thin films are also included for comparison.[20,32,39,40,42,45-49] **d** Temperature dependent cross-plane κ of a 1.75-μm-thick 3C-SiC thin film. The temperature dependent cross-plane κ of AlN and GaN thin films are also included.[20,48]

The cross-plane κ of the 3C-SiC thin films are measured by TDTR. The dependence of cross-plane κ on film thickness and temperature are shown in Fig. 4c and 4d. The cross-plane κ of 3C-SiC thin films are among the highest values, even higher than or comparable to that of diamond thin films with equivalent thickness. The cross-plane κ of 1.75-μm-thick 3C-SiC reaches ~80% of the κ of bulk 3C-SiC, up to twice as high as the κ of bulk GaN. Even the 0.93- μm-thick 3C-SiC has a cross-plane κ close to that of bulk GaN, which further confirming the high-quality of our 3C-SiC thin films. Fig. 4d compares the temperature dependent cross-plane κ of some wide bandgap semiconductor thin films. In the measured temperature range, all the cross-plane κ values of 3C-SiC are higher than that of AlN and GaN with even thicker films. The high cross-plane κ, combined with the high in-plane κ, of these 3C-SiC thin films make them the best candidate for thermal management applications which use thin films.

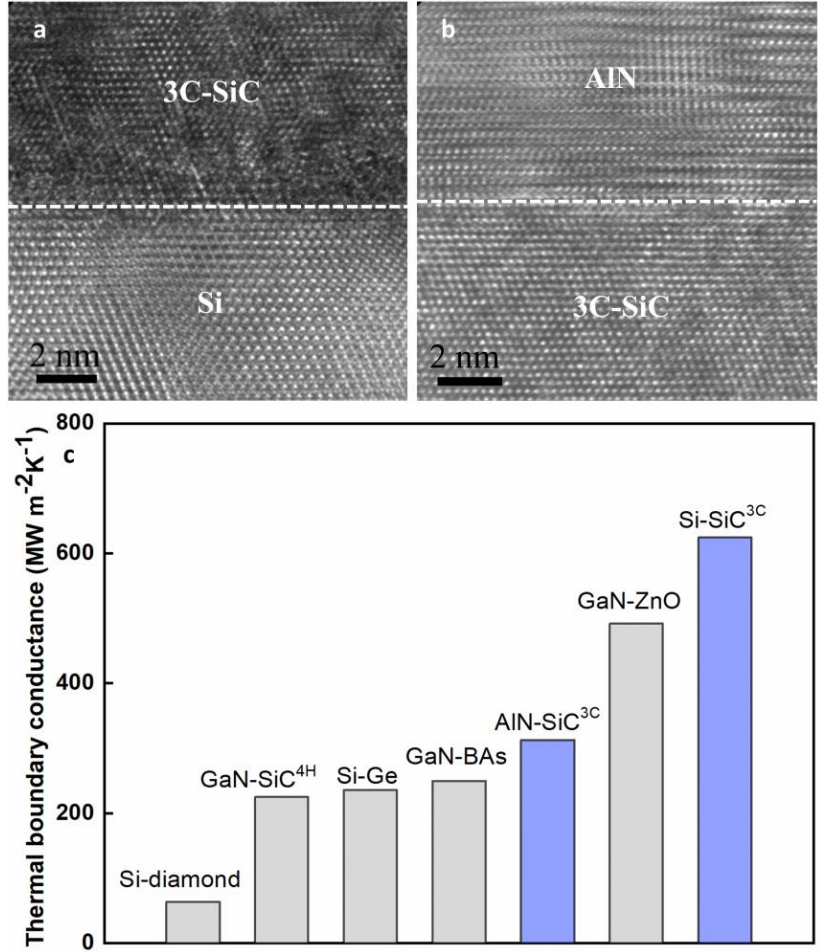

**Fig. 5. High thermal boundary conductance of 3C-SiC epitaxial interfaces. a** TEM cross-section image of Si-SiC interfaces. **b** TEM cross-section image of AlN-SiC interfaces. **c** Thermal boundary conductance of Si-SiC interfaces and AlN-SiC interfaces. The TBC values of other semiconductor interfaces are included for comparison.[28,46,49-52]

The epitaxial growth of 3C-SiC not only produces high-quality thin films which have high in-plane and cross-plane $\kappa$, but also creates high-quality heterogeneous interfaces which are potentially thermally conductance. The cross-section TEM images of the epitaxial 3C-SiC-Si and 3C-SiC-AlN interfaces are shown in Figs. 5a and 5b to study the interfacial structure. Their thermal boundary conductance (TBC) are measured by TDTR and compared with that of other semiconductor interfaces (see Fig. 5c). The measured 3C-SiC-Si TBC (~600 MW m$^{-2}$K$^{-1}$) is among the highest values for all interfaces making up of semiconductors,[51] about ten times as high as that of the diamond-Si interfaces,[46] about 2.5 times as high as that of epitaxial Si-Ge interfaces.[52] It also approaches the maximum TBC of any interface involving Si, which is only limited by the rate that thermal energy in the Si can impinge on the crystallographic plane.[53] The measured 3C-SiC-AlN TBC is higher than the GaN-BAs TBC and 4H-SiC-GaN TBC.[28,49,50] These high TBC values of 3C-SiC related interfaces facilitate heat dissipation of electronics and optoelectronics which use 3C-SiC, especially for the cases with an increasing number of interfaces as the minimization of devices.

## CONCLUSIONS

This work reported an isotropic high thermal conductivity over 500 W m$^{-1}$K$^{-1}$ at room temperature in high-purity wafer-scale free-standing 3C-SiC bulk crystals, which is ~50% higher than commercially-available 6H-SiC and AlN. It is more than 50% higher than the previously measured κ of 3C-SiC in the literature, and is the second highest among large crystals (only surpassed by single crystal diamond). We also studied the κ of corresponding 3C-SiC thin films and found record-high in-plane and cross-plane κ, even higher than diamond thin films with equivalent thickness. The measured higher κ of 3C-SiC than that of the structurally more complex 6H-SiC validates that structural complexity and κ are inversely related, resolving a long-lasting puzzle about the perplexingly low κ of 3C-SiC in the literature. Impurity concentrations measured by SIMS revealed the high-purity of our 3C-SiC crystals and the XRD measurements revealed the good crystal quality of our 3C-SiC crystals. Both contribute to the observed high κ. Furthermore, high TBC values were observed across epitaxial 3C-SiC-Si and 3C-SiC-AlN interfaces. The measured 3C-SiC-Si TBC is among the highest for semiconductor interfaces, about ten time as large as that of diamond-Si interfaces. The high κ observed in 3C-SiC bulk crystals and thin films, combined with the high TBC of epitaxial 3C-SiC interfaces, suggest 3C-SiC an excellent candidate for applications of power electronics and optoelectronics.

**METHODS**

**Sample growth**. The 3C-SiC crystals are grown on (111) Si substrates by low-temperature chemical vapor deposition (LT-CVD) in a custom-built CVD reactor at 1300 K. Because of the relatively low temperature growth, the strain in 3C-SiC layer caused by mismatch of thermal expansion coefficient is small, which enables the growth of large-diameter substrate without cracks. Since both Si and 3C-SiC have equal rotational symmetry (120º) with respect to the [111] axis, 3C-SiC crystals with less double positioning boundary (DPB) can be grown on Si. The free-standing bulk 3C-SiC crystal is obtained by growing 100-μm-thick 3C-SiC on Si substrates and then etching away the Si substrates by HNA (HF: $HNO_3$: $H_2O$). The stacking faults density observed on the growth face is about 1000 $cm^{-2}$.

**Thermal characterizations.** The κ and TBC are measured by time-domain thermoreflectance (TDTR). We usually coat 100-nm-thick Al on the to-be-measured sample as TDTR transducer before TDTR measurements. TDTR is an ultra-fast laser based pump-probe technique which can measure thermal properties of both bulk and nanostructured materials.[33,54] A modulated pump laser beam heats the sample surface periodically while a delayed probe laser beam detects the temperature variations of the sample surface via thermoreflectance. The signal picked up by a photodetector and a Lock-in amplifier is fitted with the analytical heat transfer solution of the sample structure to infer the unknown parameters (for example, κ of 3C-SiC and TBC of the metal transducer-SiC interface when measuring the 3C-SiC bulk crystals). We used 5X objective (spot size 10.7 μm) and 9.3 MHz when measuring the κ of the 3C-SiC bulk crystals and the cross-plane κ of 3C-SiC thin films. When measuring the 3C-SiC thin films, the thicknesses of Al transducer and 3C-SiC thin films are measured by picosecond acoustic technique.[55] More details about the thickness measurements and used literature values of heat capacity can be found in SI. The in-plane κ of 3C-SiC thin films are measured by BO-TDTR with a modulation frequency of 1.9 MHz and an objective of 10X.[37,38] We also used the 5X objective to repeat the BO-TDTR measurements and obtained consistent results.

**Raman spectroscopy**: Raman measurements were performed on the 3C-SiC bulk crystal with a Horiba LabRAM confocal Raman spectroscopy imaging system. The used laser wavelength is 532 nm. The acquisition time is 600 s and the objective is 50X.

**SIMS characterizations**: Secondary Ion Mass Spectrometry (SIMS) equipment (Cameca) was used to analyze the depth profile of the O, N, and B atomic densities on the C and Si faces of the 3C-SiC.

**TEM and SAED measurements:** Transmission electron microscopy (TEM) and Selected area electron diffraction (SAED) (JEM-2200FS; JEOL) were used to analyze the crystal quality of the 3C-SiC and the interface characteristics at an acceleration voltage of 200 kV. TEM samples were prepared by using a focused ion beam (FIB) system (Helios NanoLab 600i DualBeam; Thermo

Fisher Scientific) by depositing a protective layer and milling using a 30 kV accelerating voltage, and final etching step using a 2 kV accelerating voltage at room temperature.

**XRD measurements:** The crystal quality of the 3C-SiC crystals was characterized by the full-width of the half-maximum on the X-ray rocking curve of the 3C-SiC (111) peak using an X-ray diffraction system (D8 Discover; Bruker). A Cu-Kα X-ray source accelerating at 40 kV with a current of 40 mA was applied to record the XRD patterns in the range of 17.2-18.4° with a step of 0.015°. An incident slit with a width of 2 mm and a collimator with a diameter of 0.1 mm were used.

**EBSD measurements:** The crystal direction of the 3C-SiC crystals was analyzed by an Electron Backscatter diffraction (EBSD) system (FE-SEM JSM-6500F; JEOL) with a high-resolution scanning electron microscope (SEM) and a TSL orientation imaging microscopy (OIM) analyzer. The SEM was operated at 20 kV, and a scan area of 2.4 mm×0.8 mm was performed using a hexagonal grid with a step size of 2 μm.

## DATA AVAILABILITY

The datasets generated during and/or analyzed during the current study are available from the corresponding authors upon reasonable request.

## CODE AVAILABILITY

The code used for calculations, simulations, and data analysis is available from the corresponding authors upon reasonable request.

## ACKNOWLEDGEMENTS

Z.C. S.G. and D.C. would like to acknowledge the financial support from U.S. Office of Naval Research under a MURI program (Grant N00014-18-1-2429). Z.C. and D.C. would like to Guangxin Lyu for help of Raman measurements. The fabrication of the TEM samples was performed at The Oarai Center and at the Laboratory of Alpha-Ray Emitters in IMR under the Inter-University Cooperative Research in IMR of Tohoku University (NO. 202112-IRKMA-0016). The observation of the TEM samples was supported by Kyoto University Nano Technology Hub in the "Nanotechnology Platform Project" sponsored by the Ministry of Education, Culture, Sports, Science and Technology (MEXT), Japan.

## AUTHOR CONTRIBUTIONS

Z.C. initialized this project and finished all the thermal measurements. Z.C. and J.L. coordinated the project. J.L. polished the bulk 3C-SiC samples and performed TEM studies. K.K., H.A., and H.U. grew the 3C-SiC samples and did XRD, EBSD, and SIMS measurements. Y.O. and Y.N. prepared the TEM samples. Z.C. wrote the manuscript with inputs from all authors. J.L. assisted with manuscript preparation and reviewed the manuscript. S.G. and N.S. commented on the manuscript. D.C. provided overall guidance to the project and reviewed the manuscript.

## COMPETING INTERESTS

The authors declare no competing interest.

## SUPPLEMENTARY INFORMATION

Additional details are provided in the SI, including details about sample characterizations, details about TDTR measurements, and details about BO-TDTR measurements.